\documentclass[useAMS]{mn2e}
\usepackage{times}
\usepackage{graphicx}
\usepackage{amsmath}

\title[Relativistic Fermi acceleration with shock compressed
turbulence]{Relativistic Fermi acceleration with shock compressed turbulence}

\author[M. Lemoine and B. Revenu]{Martin Lemoine$^{1}$\thanks{E-mail:
{\tt lemoine@iap.fr}} and Beno\^\i t
Revenu$^{2,1}$\thanks{E-mail: {\tt revenu@iap.fr}}\\
$^1$ Institut d'Astrophysique de Paris, \\ UMR 7095 CNRS, Universit\'e
Pierre \& Marie Curie\\ 98 bis boulevard Arago, F-75014 Paris,
France\\ 
$^2$ AstroParticule \& Cosmologie, \\ UMR 7164 CNRS, Universit\'e
Denis Diderot}

\begin{document}

\date{}

\pagerange{\pageref{firstpage}--\pageref{lastpage}} \pubyear{2005}

\maketitle

\label{firstpage}

\begin{abstract}
  This paper presents numerical simulations of test particle Fermi
  acceleration at relativistic shocks of Lorentz factor $\Gamma_{\rm
  sh}=2-60$, using a realistic downstream magnetic structure obtained
  from the shock jump conditions. The upstream magnetic field is
  described as pure Kolmogorov turbulence; the corresponding
  downstream magnetic field lies predominantly in the plane tangential
  to the shock surface and the coherence length is smaller along the
  shock normal than in the tangential plane.  Acceleration is
  nonetheless efficient and leads to powerlaw spectra with index
  $\simeq 2.6-2.7$ at large shock Lorentz factor $\Gamma_{\rm sh}\gg1$,
  markedly steeper than for isotropic scattering downstream. The
  acceleration timescale $t_{\rm acc}$ in the upstream rest frame
  becomes a fraction of Larmor time $t_{\rm L}$ in the
  ultra-relativistic limit, $t_{\rm acc}\approx 10 t_{\rm
  L}/\Gamma_{\rm sh}$. Astrophysical applications are discussed, in
  particular the acceleration in $\gamma-$ray bursts internal and
  external shocks.
\end{abstract}


\section{Introduction}

The Fermi acceleration process of charged particles bouncing back and
forth across a shock wave is the main ingredient for the generation of
high energy radiation in a variety of astrophysical environments. This
observed radiation is generally synchrotron light emitted by the
accelerated electrons; in this case one may recover the spectral index
$s$ of the accelerated population from the synchrotron index.  For
example, the afterglow emission of $\gamma-$ray bursts that is seen in
X-ray through the infrared is generally interpreted as synchrotron
emission of electrons accelerated at the ultra-relativistic external
shock of Lorentz factor $\Gamma_{\rm sh}\sim 300$. The inferred
spectral index, $s\simeq 2.3\pm0.1$ (Waxman 1997; see also Meszaros
2002 and Piran 2004 for reviews), thus probes the nature of shock
acceleration in the ultra-relativistic regime.  Similarly synchrotron
emission of electrons accelerated in the mildly relativistic internal
shocks ($\Gamma_{\rm sh}\sim\,2-5$ in the comoving frame) with index
$s\simeq 2.3\pm0.1$ could explain the prompt $\gamma$ emission (see
e.g. Meszaros 2002, Piran 2004 and references therein). These
observations thus provide anchor points for studies of Fermi
acceleration in the moderate to the ultra- relativistic regime. They
have actually been regarded as a dramatic confirmation of the theory
of shock acceleration in the relativistic regime, which has been
claimed to predict a ``universal'' asymptotic spectral index $s \simeq
2.23$ in the ultra-relativistic regime $\Gamma_{\rm sh}\gg 1$.

Relativistic shock acceleration has been studied through a variety of
methods, either analytical (Peacock 1981, and more recently Vietri
2002, Vietri 2003, Keshet \& Waxman 2005, Blasi \& Vietri 2005),
semi-analytical (Kirk \& Schneider 1987; Gallant \& Achterberg 1999;
Kirk {\it et al.} 2000; Achterberg {\it et al.} 2001), or numerical
(Ellison {\it et al.} 1990, Ostrowski 1991, Ballard \& Heavens 1992,
Ostrowski 1993, Bednarz \& Ostrowski 1996, 1998, 1999, Ellison \&
Double 2002, 2004, Lemoine \& Pelletier 2003, Meli \& Quenby 2003a,b,
Bednarz 2004, Niemec \& Ostrowski 2004, Baring 2004). Not all of these
studies find the universal value for $s$, however, all the more so
when anisotropic configurations such as oblique shocks are
considered. One clear example is the demonstration that Fermi
acceleration in superluminal (perpendicular) shocks in the absence of
cross-field diffusion becomes inefficient (Begelman \& Kirk 1990); in
the relativistic regime, oblique shocks are superluminal unless the
angle between the magnetic field and the shock normal $\Theta_{B}\la
1/\Gamma_{\rm sh}$.

It is generally suspected that the inclusion of scattering would make
Fermi acceleration more efficient in the relativistic regime. The
simulations of Bednarz \& Ostrowski (1998) and Baring (2004) have
indeed confirmed that the spectral slope tends to increase with
increasing shock obliquity and with decreasing turbulence level, whose
r\^ole is to permit cross-field line transport to the shock
front. However, a limitation of these simulations is that the
scattering is simulated in a phenomenological way by setting a ratio
of the perpendicular to parallel diffusion lengths and drawing pitch
angles at random at each time step. Upstream, it has been demonstrated
that an ultra-relativistic shock wave overtakes the particle before
this latter has had time to scatter efficiently (Gallant \& Achterberg
1999, Achterberg {\it et al.} 2001), so that the details of particle
transport are probably not crucial.  Downstream, however, the particle
has to turn back before re-crossing the shock, and the approximation
of ad-hoc diffusion lengths may be too na\"\i ve to accurately
simulate the transport. In fact, one may expect non-trivial
correlation functions between displacements along different directions
as well as subdiffusion regimes to play a significant r\^ole in the
return to the shock.

Several studies have tried to integrate out exactly the particles
trajectories in a well-defined magnetic field structure (Ballard \&
Heavens 1992, Ostrowski 1993, Lemoine \& Pelletier 2003, Niemec \&
Ostrowski 2004). The study of Ballard \& Heavens (1992) involved a
realistic magnetic field structure, in the sense that it obeys the
shock jump conditions, but was limited to mildly relativistic shocks
($\Gamma_{\rm sh}\leq 5$). Nevertheless, it observed a trend of
increasing spectral index with increasing shock velocity, a result
which has been disputed by the more exhaustive simulations of
Ostrowski (1993). The recent work of Niemec \& Ostrowski (2004)
considered situations of moderate turbulence levels with varying
degrees of obliquity for mildly relativistic shocks ($\Gamma_{\rm
sh}\leq 5$). The conclusions obtained indicate that various spectral
slopes can indeed be obtained, although the noise on the simulations
is not negligible. Finally, the work of Lemoine \& Pelletier (2003)
introduced a new numerical Monte Carlo method to study relativistic
Fermi acceleration, on which the present work is based. It assumed the
downstream turbulence to be isotropic and confirmed the value $s\simeq
2.2-2.3$ predicted in that case.

It seems fair to say that a clear picture of the efficiency of
relativistic Fermi acceleration in a magnetic structure that includes
compression of the upstream magnetized configuration has not yet
emerged. The present paper proposes to undertake such simulations in
order to make progress along these lines. We assume that the upstream
magnetic field is described by pure Kolmogorov turbulence, i.e. there
is not uniform component; this can be seen as the limit $\delta B/B
\rightarrow \infty$ of a highly turbulent plasma. The upstream
magnetic field is compressed by the shock into an anisotropic
downstream turbulence. We conduct our simulations in the mildly and
ultra-relativistic regimes; simulations of that kind in this latter
regime have never been attempted before.

In Section~2, we describe in detail the numerical techniques and the
procedure used (borrowed from Lemoine \& Pelletier 2003) to simulate
the Fermi acceleration process. In Section~3, we present our results
on the (downstream) return probability, the acceleration timescale and
the accelerated spectrum as a function of shock Lorentz factor. In
Section~4 we discuss the relaxation length of the turbulence and argue
that, for relativistic shocks at least, particles that return to the
shock downstream do not travel beyond the point where the anisotropy
of turbulence has relaxed. We also discuss the properties of transport
of particles in the strongly anisotropic turbulence generated by shock
compression, compare our results to previous studies and comment on
the applications of our results to shock acceleration in $\gamma-$ray
bursts and to shock acceleration of ultra-high energy cosmic rays.
Conclusions and a summary of the results are provided in Section~5.

\section{Numerical simulations}\label{numrec}

\subsection{Jump conditions and magnetic fields}

In the present work, we assume that the magnetic field is dynamically
unimportant, {\it i.e.} its energy density can be neglected with
respect to that of the fluid. We also consider a strong shock, for
which the upstream random kinetic energy per particle can be neglected
with respect to that downstream.  The corresponding hydrodynamic jump
conditions are given in Blandford \& McKee (1977), and reviewed in
Kirk \& Duffy (2001) and Gallant (2002).  The shock Lorentz factor is
denoted $\Gamma_{\rm sh}$ in the upstream frame (taken as the lab
frame), and the shock velocity upstream is $\beta_{\rm sh}$. Unless
otherwise noted, all quantities are calculated in this frame. If
relevant the reference frame is indicated by a subscript, e.g.,
$\beta_{\rm sh\vert d}$ refers to the shock velocity measured in the
downstream rest frame and $\Gamma_{\rm sh\vert d}$ refers to the shock
Lorentz factor in the downstream frame. The relative Lorentz factor
between upstream and downstream is noted $\Gamma_{\rm rel}$ and reads:
$$\Gamma_{\rm rel}\equiv \Gamma_{\rm sh} \Gamma_{\rm sh\vert
d}(1-\beta_{\rm sh}\beta_{\rm sh\vert d}).$$ The downstream Lorentz
factor $\Gamma_{\rm sh\vert d}$ as well as $\Gamma_{\rm rel}$ can be
obtained as a function of $\Gamma_{\rm sh}$ (upstream shock Lorentz
factor) using the relations derived from the shock jump conditions
for a Synge equation of state (Gallant 2002):
\begin{equation}
  \Gamma_{\rm sh\vert d}^2 \,  =  \, \frac{F(\xi)}{F(\xi)-1}\, , \quad
  \Gamma_{\rm sh}^2 \,  =  \, G(\xi)^2 \frac{F(\xi)}{F(\xi)-1}\
     \label{eq:BDE}
\end{equation}
where $\xi \equiv mc^{2}/T_{\rm d}$, $T_{\rm d}$ being the downstream
temperature and $m$ the particle mass, $G(\xi) \equiv
K_{3}(\xi)/K_{2}(\xi)$, with $K_2$, $K_3$ modified Bessel functions,
and $F(\xi)\equiv [\xi G(\xi) - 1]^2 - \xi^2$. These relations hold
for a gas composed of possibly different particles species but with
same $\xi$ (Gallant 2002). Equations~(\ref{eq:BDE}) can be inverted
numerically to obtain $\Gamma_{\rm sh \vert d}$ as a function of
$\Gamma_{\rm sh}$.  In particular, in the ultra-relativistic limit
$\Gamma_{\rm sh}\rightarrow +\infty$, one finds the well-known results
$\beta_{\rm sh\vert d}\rightarrow 1/3$ ($\Gamma_{\rm sh\vert
d}\rightarrow 3/\sqrt{8}$) and $\Gamma_{\rm rel}\rightarrow
\Gamma_{\rm sh}/\sqrt{2}$.

  The conservation of the electromagnetic field energy-momentum tensor
implies the following jump conditions for the magnetic field
components $B_\parallel$ (aligned with the shock normal) and $B_\perp$
(tangential to the shock surface):

\begin{equation}
{B_{\parallel, \rm d \vert d}\over B_{\parallel, \rm u\vert
u}}\,=\,1,\quad\quad {B_{\perp, \rm d\vert d}\over B_{\perp, \rm u \vert
u}}\,=\, {\beta_{\rm sh\vert u}\Gamma_{\rm sh\vert u} \over
\beta_{\rm sh\vert d}\Gamma_{\rm sh\vert d}},
\end{equation}
and as before, $\beta_{\rm sh \vert u} \equiv \beta_{\rm sh}$,
$\Gamma_{\rm sh \vert u} \equiv \Gamma_{\rm sh}$. The parallel
component $B_\parallel$ is thus conserved while the perpendicular
component $B_\perp$ is amplified by the proper shock compression ratio
$R=\beta_{\rm sh\vert u}\Gamma_{\rm sh\vert u} / \beta_{\rm sh\vert
d}\Gamma_{\rm sh\vert d}$. In the ultra-relativistic limit
$R\rightarrow \Gamma_{\rm sh}\sqrt{8}$, and the total magnetic field
strength is amplified by $\sqrt{2/3}R$.

We assume that the upstream magnetic field is purely turbulent with a
power spectrum describing Kolmogorov turbulence with maximal length
scale $L_{\rm max}$. It is modeled as a sum of static plane wave modes
according to:
\begin{equation}
{\mathbf B}_{\rm u}({\mathbf x}) \,=\, \sum_{\mathbf k}\, e^{i{\mathbf
k}\cdot{\mathbf x}+i\phi_{\mathbf k}}\,{\mathbf e}_{\mathbf k}G_k,\label{eq:Bu}
\end{equation}
with $\mathbf{e}_{\mathbf k}$ a unit polarization vector orthogonal to
$\mathbf k$, $\phi_{\mathbf k}$ a random phase and $\vert G_k\vert^2
\propto k^{-5/3}$ the amplitude of the power spectrum. The wavenumbers
$k$ range from $k_{\rm min}=2\pi/L_{\rm max}$ to some maximal
wavenumber $k_{\rm max}\gg k_{\rm min}$; numerically we employ~250
wavenumbers modes whose directions are drawn at random, and whose
moduli are spaced logarithmically between $k_{\rm min}$ and $k_{\rm
  max} = 5\cdot10^3 k_{\rm min}$. The amplitude $G_k$ can be chosen as
real and is normalized such that:
\begin{equation}
{1\over V}\int {\rm d}{\mathbf x}\, {\mathbf B}^2({\mathbf x}) \, = \, \sum_{\mathbf
k} \vert G_k\vert^2\,\equiv\, B_{\rm rms}^2
\end{equation}
with $B_{\rm rms}^2$ the squared turbulent magnetic field strength.

According to the shock jump conditions, the downstream magnetic field
is described by an anisotropic turbulence: while $B_\parallel$ is
conserved, the turbulence wavenumbers $k_\parallel$ are amplified by
$R$, which corresponds to the compression of the eddies by $1/R$ along
the shock normal. The perpendicular wavenumbers $k_\perp$ are
conserved through the shock but $B_\perp$ is amplified as before.
Hence the downstream magnetic field is described by:
\begin{equation}
{\mathbf B}_{\rm d}({\mathbf x}) \,=\, \sum_{\mathbf{ \tilde k}}\,
e^{i{\mathbf {\tilde k}}\cdot{\mathbf x}+i\phi_{\mathbf{\tilde
k}}}\,{\mathbf{ \tilde e}}_{\mathbf{\tilde k}}G_{k},
\end{equation}
where $\mathbf{ \tilde k}$ is related to the wavenumber $\mathbf k$ of
Eq.~\ref{eq:Bu} by ${\mathbf{ \tilde k}}_{\parallel} = R\, {\mathbf
k}_{\parallel}$ and ${\mathbf{ \tilde k}}_{\perp} = {\mathbf
k}_{\perp}$; similarly ${\mathbf{ \tilde e}}_{\mathbf k}$ is related
to ${\mathbf e}_{\mathbf k}$ by: ${\mathbf{ \tilde e}}_{\parallel,
\mathbf k} = {\mathbf e}_{\parallel, \mathbf k}$ and ${\mathbf{ \tilde
e}}_{\perp,\mathbf k} = R\, {\mathbf e}_{\perp,\mathbf k}$. Note that
$\mathbf{\tilde k}\cdot\mathbf{\tilde e}_{\mathbf{\tilde
k}}=\mathbf{k}\cdot\mathbf{e}_{\mathbf k}=0$ as required for a
divergenceless field; $\phi_\mathbf{\tilde k}$ and $G_{k}$ are not
modified. We chose not to normalize the above polarization vector to
unity downstream and its modulus gives the overall amplification
factor of the magnetic field. This is but a matter of convention: one
may equally well embody the compression factor in $G_k$ and normalize
$\mathbf{\tilde e}_{\mathbf{\tilde k}}$ to unity.

\subsection{Monte Carlo simulations}

Our numerical procedure is summarized in Lemoine \& Pelletier
(2003). It consists in two main steps: in a first stage, we conduct
Monte Carlo simulations of particle propagation in a magnetized medium
(either upstream or downstream) and derive the statistical properties
related to shock crossing and re-crossing, as described below. In a
second step we use these statistical distributions in conjunction with
the Lorentz transforms from one frame to the other to reconstruct the
accelerated spectrum that escapes downstream.

 Once the magnetic field structure is set up as described above, one
Monte Carlo simulation of the propagation of particle consists in
integrating the equation of motion in the magnetic field. The particle
trajectory is saved in time intervals that are a fraction $f_{\rm
u}\simeq 10^{-4}$ (upstream) or $f_{\rm d}\simeq 10^{-2}$ (downstream)
of Larmor time $t_{\rm L}=R_{\rm L}/c$ (with $R_{ rm L}\equiv p/qB$)
over a time period as long as $\Delta T_{\rm u}\simeq 10^{2}$
(upstream) or $\Delta T_{\rm d}\simeq 10^4$ (downstream) Larmor time.
For each computed trajectory one can build a statistical sample of
shock crossing and re-crossing as follows. One draws at random a point
along the trajectory, which defines the point at which the particle
enters through the shock. One records the ingress pitch angle cosine
of the particle momentum with respect to the shock normal at that
point. One then searches for the point along the trajectory at which
the particle exits through the shock and the corresponding egress
pitch angle cosine is recorded. In the downstream medium, it happens
that the particle never re-crosses the shock as the shock itself moves
away with speed $\beta_{\rm sh\vert d}\simeq 1/3$ ($\Gamma_{\rm
sh}\gg1$). By iterating the above procedure, {\it i.e.}  drawing other
points of entry in the trajectory, and building other trajectories,
one can measure the probability laws that control in a direct manner
the Fermi process.

In particular, the ratio of the number of shock re-crossings to the
total number of shock entries at a given ingress ``pitch'' angle
(defined here as the angle between the momentum and the direction of
the shock normal) cosine $\mu^{\rm i}$ gives the return probability
$P_{\rm ret}(\mu^{\rm i})$.  In a similar way, the number of shock
re-crossings through an egress pitch angle cosine $\mu^{\rm e}$ for a
given ingress cosine $\mu^{\rm i}$ gives (after proper normalization)
the conditional return probability ${\cal P}(\mu^{\rm i};\mu^{\rm
e})$. One can define and calculate these quantities both downstream,
${\cal P}_{\rm d}(\mu^{\rm i}_{\rm d};\mu^{\rm e}_{\rm d})$, and
upstream, ${\cal P}_{\rm u}(\mu^{\rm i}_{\rm u};\mu^{\rm e}_{\rm
u})$. Note that the ingress and egress pitch angles are calculated in
the rest frame of the fluid under consideration.  The normalization of
the conditional probability laws is such that their sum over the
egress pitch angle cosine yields the return probability as a function
of ingress pitch angle cosine:
\begin{eqnarray}
P_{\rm ret,\,d}(\mu^{\rm i}_{\rm d})\,&=&\, \int {\rm d}\mu^{\rm e}_{\rm
d}\, {\cal P}_{\rm d}(\mu^{\rm i}_{\rm d};\mu^{\rm e}_{\rm
d}),\nonumber\\
P_{\rm ret,\,u}(\mu^{\rm i}_{\rm u})\,&=&\, \int {\rm
d}\mu^{\rm e}_{\rm u}\, {\cal P}_{\rm u}(\mu^{\rm i}_{\rm u};\mu^{\rm
e}_{\rm u}).
\label{eq:pret}
\end{eqnarray}

  Obviously the upstream return probability $P_{\rm ret,\,u}$ must be
unity if one considers an infinite planar shock with an infinite
lifetime. This provides a useful check on the numerical procedure; in
the present calculations, $P_{\rm ret,\,u}$ does not deviate from
unity by more than $\sim 10^{-6}$. Downstream it is mandatory to
verify that one does not miss possible late returns by varying the
trajectory integration time; we estimate that the mean of $P_{\rm
ret,\,d}$ over ingress pitch angle cosines is accurate to better than
$\sim 10^{-4}$.

 Finally these simulations give a direct measurement of the return
timescale to the shock as a function of pitch angles. This measurement
is particularly important to estimate the maximal acceleration energy
in a variety of environments, as discussed in Section~3.3.

Once the upstream and downstream laws of return probability are known,
the simulation of the acceleration process itself can be performed as
follows. We denote by ${\cal F}_{\rm d}^{2n+1}(\mu_{\rm
  d},\epsilon_{\rm d})$ the distribution function of particles that
enter the shock towards downstream with ingress pitch angle cosine
$\mu_{\rm d}$, that have experienced $2n+1$ shock crossings and that
carry energy $\epsilon_{\rm d}$ (downstream frame).  Similarly we
define the distribution function ${\cal F}_{\rm u}^{2n}(\mu_{\rm
  u},\epsilon_{\rm u})$ of upstream-going particles with ingress pitch
angle cosine $\mu_{\rm u}$, having experienced $2n$ shock crossings
and carrying energy $\epsilon_{\rm u}$. If we denote by ${\cal F}_{\rm
  u}^0$ the injection population upstream, then after an even (resp.
odd) number of shock crossings the particles are necessarily upstream
(resp. downstream). The injection (isotropic) distribution
function ${\cal F}_{\rm u}^0$ is normalized to unity, as follows:
\begin{equation}
\int_{-1}^1{\rm d}\mu_{\rm u}{\rm d}\epsilon_{\rm u}\,{\cal F}^{0}_{\rm
u}(\mu_{\rm u},\epsilon_{\rm u})\,\equiv\,1.
\label{eq:norm}
\end{equation}
The integral over $\mu$ and $\epsilon$ of the distribution functions
${\cal F}_{\rm d}^{2n+1}(\mu,\epsilon)$ and 
${\cal F}_{\rm u}^{2n}(\mu,\epsilon)$ with $n>0$ is smaller than
unity, due to escape of particles downstream at each cycle.

 Now, particles that enter upstream after $2n$ shock crossings with
ingress cosine $\mu_{\rm u}^{\rm i}$ re-cross the shock with egress
cosine $\mu_{\rm u}^{\rm e}$ and with conditional probability ${\cal
P}_{\rm u}(\mu_{\rm u}^{\rm i};\mu_{\rm u}^{\rm e})$. The total number
of particles with egress pitch angle $\mu_{\rm u}^{\rm e}$ and energy
$\epsilon_{\rm u}$ at the $2n+1^{\rm th}$ shock crossing is $\int {\rm
d}\mu_{\rm u}^{\rm i}\,{\cal P}_{\rm u}(\mu_{\rm u}^{\rm i};\mu_{\rm
u}^{\rm e}){\cal F}_{\rm u}^{2n}(\mu_{\rm u}^{\rm i},\epsilon_{\rm u})$. We
note that the upstream egress cosine $\mu_{\rm u}^{\rm e}$ is related
to the corresponding downstream ingress cosine $\mu_{\rm d}^{\rm i}$,
by a Lorentz transform, just as the energies measured in the upstream
frame ($\epsilon_{\rm u}$) or downstream frame ($\epsilon_{\rm d}$):
\begin{equation}
\mu_{\rm d}^{\rm i} \, = \, \frac{\mu_{\rm u}^{\rm e} - \beta_{\rm
rel}}{1-\beta_{\rm rel}\mu_{\rm u}^{\rm e}} ,\quad \epsilon_{\rm d} \, =
\, \Gamma_{\rm rel}(1-\beta_{\rm rel}\mu^{\rm e}_{\rm u})\epsilon_{\rm u}
,
\label{eq:mapud_L}
\end{equation}
with a similar relation between $\mu_{\rm d}^{\rm e}$ and $\mu_{\rm
u}^{\rm i}$ when the particle crosses the shock from downstream to
upstream.

Therefore, the conservation of particle number at shock crossing
u$\rightarrow$d implies the following relation between ${\cal F}_{\rm
d}^{2n+1}$ and ${\cal F}_{\rm u}^{2n}$:

\begin{equation}        
\begin{split}   
{\cal F}_{\rm d}^{2n+1}(\mu_{\rm d}^{\rm i},\epsilon_{\rm d})\,{\rm
d}\mu_{\rm d}^{\rm i}{\rm d}\epsilon_{\rm d} =\quad\quad\quad\quad\quad\quad\quad\quad\quad\quad\quad\quad \\
\, \left[ \int_{\beta_{\rm sh}}^{1}{\rm d}\mu_{\rm u}^{\rm i}\,{\cal
P}_{\rm u}(\mu_{\rm u}^{\rm i}; \mu_{\rm u}^{\rm e}) {\cal F}_{\rm
u}^{2n}(\mu_{\rm u}^{\rm i},\epsilon_{\rm u})\right]
{\rm d}\mu_{\rm u}^{\rm e}{\rm d}\epsilon_{\rm u},
\label{eq:mapud_f}
\end{split}
\end{equation}
and one obtains a similar system for shock crossing~d$\rightarrow$u:\\
\begin{equation}
\begin{split}
{\cal F}_{\rm u}^{2n}(\mu_{\rm u}^{i},\epsilon_{\rm u}) \,{\rm d}\mu_{\rm u}^{\rm i}{\rm
d}\epsilon_{\rm u} = \quad\quad\quad\quad\quad\quad\quad\quad\quad\quad\quad\quad\quad\quad\\ \, 
\left[\int_{-1}^{\beta_{\rm sh\vert d}}{\rm d}\tilde\mu_{\rm d}^{\rm
i}\,{\cal P}_{\rm d}(\tilde\mu_{\rm d}^{\rm i}; \mu_{\rm d}^{\rm e})
{\cal F}_{\rm d}^{2n-1}(\tilde\mu_{\rm d}^{\rm i},\tilde\epsilon_{\rm
  d})\right]
\,{\rm d}\mu_{\rm d}^{\rm e}{\rm d}\tilde\epsilon_{\rm d},
\label{eq:mapdu_f}
\end{split}
\end{equation}
with:
\begin{equation}
\mu_{\rm u}^{\rm i} \, = \, \frac{\mu_{\rm d}^{\rm e} + \beta_{\rm
rel}}{1+\beta_{\rm rel}\mu_{\rm d}^{\rm e}}, \quad \epsilon_{\rm u} \,
= \, \Gamma_{\rm rel}(1+\beta_{\rm rel}\mu^{\rm e}_{\rm
d})\tilde\epsilon_{\rm d}
\label{eq:mapdu_L}
\end{equation}
where the ``~~$\mathbf{\widetilde{~}}$~~'' symbol has been introduced
to differentiate the values of $\mu_{\rm d}$ and $\epsilon_{\rm d}$
from one cycle ($2n-1$ shock crossings) to the next ($2n+1$ shock
crossings). The integration bounds on $\mu$ are imposed by the shock
crossing conditions.

  The terms within brackets in Eqs.~(\ref{eq:mapud_f}) and
(\ref{eq:mapdu_f}) correspond to the distributions upon exit from
upstream and downstream respectively. These equations assume
implicitly that the conditional probability laws do not depend on
energy. This will be shown to be a good approximation in Section 3.1.

  After each cycle u$\rightarrow$d$\rightarrow$u, a population ${\cal
F}_{\rm out}^{2n+1}(\epsilon_{\rm d})=\int {\rm d}\mu_{\rm d}^{\rm i}
[1- P_{\rm ret}(\mu_{\rm d}^{\rm i})]{\cal F}_{\rm d}^{2n+1}(\mu_{\rm
d}^{\rm i};\epsilon_{\rm d})$ of the particle population has escaped
downstream. The sum over $n$ of these escaping particles forms the
outgoing accelerated particle population:
\begin{equation}
{\cal F}_{\rm out}(\epsilon)\,=\,\sum_{n=0}^{n=+\infty} {\cal F}_{\rm
out}^{2n+1}(\epsilon).
\label{eq:fout}
\end{equation}
By following each shock crossing, and using Eqs.~(\ref{eq:mapud_L}),
(\ref{eq:mapud_f}), (\ref{eq:mapdu_f}), (\ref{eq:mapdu_L}) and
(\ref{eq:fout}) one can follow the evolution of ${\cal F}_{\rm d}$,
${\cal F}_{\rm u}$ and ${\cal F}_{\rm out}$, starting from a
mono-energetic and isotropic initial injection distribution
upstream. A similar formal development of the acceleration process by
repeated shock crossings has also been proposed independently by
Vietri (2002): the flux of particles crossing the shock in the
stationary regime, noted $J_{\rm in}$ in Vietri (2002) is related to
the above as $J_{\rm in} = C \sum_{n=0}^{n=+\infty} {\cal F}_{\rm
d}^{2n+1}$ with $C$ a normalization constant (see also Lemoine \&
Pelletier 2003).

The present technique has significant advantages when compared to
standard Monte Carlo techniques which follow the particle trajectories
on both sides of the shock through the whole acceleration process; in
particular, it offers a significant gain in signal to noise as will be
obvious in Section 3. 

It has however one caveat that should be underlined and which resides
in the fact that we compute the accelerated spectrum by merging
separate pieces of information on transport properties upstream and
downstream. In so doing, we neglect the possible correlations that may
exist between the upstream magnetic configuration at the crossing
point and that downstream, i.e. we neglect the possible correlations
between upstream and downstream transport. The only method that can
take this effect into account is the direct Monte Carlo integration of
individual particle trajectories. It is therefore important to compare
the results obtained with these two methods in order to assess the
magnitude of this effect. In the case where scattering is isotropic
downstream, it appears that various methods converge to the same value
of the spectral index, and this includes various non-Monte Carlo
methods which cannot take the above effect into account (see, e.g.,
Achterberg {\it et al.} 2001, Lemoine \& Pelletier 2003, Keshet \&
Waxman 2005) as well as direct Monte Carlo methods (e.g., Bednarz \&
Ostrowski 1998). This suggests that, at least in the isotropic limit,
these correlations do not play a significant r\^ole in the
determination of the spectral index.

\section{Results}

The numerical technique described in the previous section allows to
collect a significant amount of information on the acceleration
process, in particular the conditional probabilities of return from
downstream or upstream, the energy gain per cycle as well as the
acceleration timescale. In order to better understand the results
obtained for each of these quantities, it is necessary to emphasize
the difference between the effective coherence length along the shock
normal $L_\parallel$ and that tangential to the shock front $L_\perp$
as measured downstream, see Section~2: $L_\parallel = L_\perp / R =
L_{\rm max}/R$, where $L_{\rm max}$ is the coherence length of the
upstream magnetic field, and $R$ the proper shock compression ratio,
$R\simeq \Gamma_{\rm sh}\sqrt{8}$ when $\Gamma_{\rm sh}\gg1$. This
distinction takes on a particular importance when one compares the
results over various values of the shock Lorentz factor and over
various values of the rigidity $\rho\equiv2\pi R_{\rm L}/L_{\rm max}$,
where $R_{\rm L}$ denotes the Larmor radius in the rest frame of
consideration. In principle, the transport properties of particles in
a magnetic field depend solely on the rigidity. However, when the
effective coherence length along the shock normal depends on
$\Gamma_{\rm sh}$ through $R$, while $L_\perp$ does not, there is no
unambiguous definition of rigidity. In particular, the above
definition of $\rho$ does not correspond to the effective rigidity
$\rho_\parallel \equiv 2\pi R_{\rm L}/L_\parallel$ that controls the
scattering of particles with turbulence modes of wavevector parallel
to the shock normal: a given rigidity $\rho=2\pi R_{\rm L}/L_{\rm
  max}$ corresponds in fact to larger and larger values of
$\rho_\parallel$ as $\Gamma_{\rm sh}$ increases. The relevance of this
observation to the results will be adressed shortly.

\subsection{Return probability}

\begin{figure}
\centering \includegraphics[width=0.5\textwidth,clip=true]{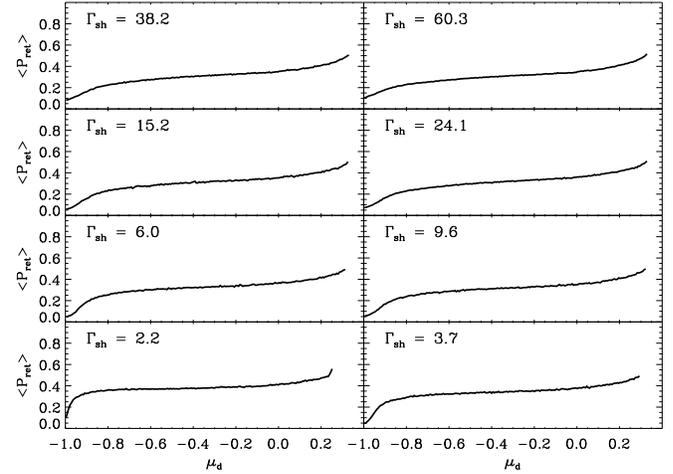}
\caption[]{Downstream return probability {\it vs} ingress pitch angle cosine
$\mu_d^i$ (downstream rest frame) for various shock Lorentz factors, as
indicated, and for a rigidity $2\pi R_{\rm L}/L_{\rm max}=6\cdot 10^{-4}$.}
\label{fig:Pret_mu}
\end{figure}

The average return probability marginalized over egress angle, $P_{\rm
ret, d}$, defined in Eq.~\ref{eq:pret} as the direct average of the
conditional probability law ${\cal P}_{\rm d}(\mu^{\rm i}_{\rm
d},\mu^{\rm e}_{\rm d})$ over the egress pitch angle cosine $\mu^{\rm
e}_{\rm d}$, is shown as a function of the ingress pitch angle cosine
$\mu^{\rm i}_{\rm d}$ in Fig.~\ref{fig:Pret_mu}. The increase of
$P_{\rm ret, d}$ as $\mu^{\rm i}_{\rm d}\to 1/3$ backs up the notion
that particles crossing the shock from upstream to downstream at near
grazing incidence with the shock front have a substantially higher
probability of returning to the shock than those crossing the shock
head-on ($\mu^{\rm i}_{\rm d}\to -1$).

  When viewed as a function of shock Lorentz factor, the average
return probability appears to reach an asymptotic law, as is made
apparent in Fig.~\ref{fig:Pret_mu}. This is not a trivial result in
itself, as the nature of the turbulence downstream depends rather
strongly on the shock Lorentz factor. Section~4.1 provides examples of
downstream trajectories for two different values of $\Gamma_{\rm sh}$,
and indeed, the displacements along the shock normal differ
widely. Hence one might naturally expect that the return probability
would carry some form of dependence on the shock Lorentz factor
$\Gamma_{\rm sh}$. As we now argue, this is related to the fact that
the scattering timescale in the direction along the shock normal,
i.e. the time required for the particle to turn back, is a function of
Larmor time, as demonstrated in Section~4.1. There it is argued that
the particles that return to the shock have done one reflection on the
compressed turbulence in their first interaction; indeed, particles
get trapped in a layer of the compressed turbulence when $\Gamma_{\rm
sh}\gg1 $ and $\rho\ll1$, hence they cannot return to the shock unless
they do so in the first interaction. This reflection is in fact a
half-gyration of the particle around a field line which is mainly
oriented along the shock front as a result of shock compression, and
this explains why the scattering time is of order of the Larmor
time. For large values of $\Gamma_{\rm sh}$, the shock velocity with
respect to downstream $\beta_{\rm sh\vert d}\rightarrow 1/3$ becomes
independent of $\Gamma_{\rm sh}$, and so does the scattering timescale
(at a given rigidity, see Section 4.1), hence so does the return
probability.

\begin{figure}
\centering \includegraphics[width=0.5\textwidth,clip=true]{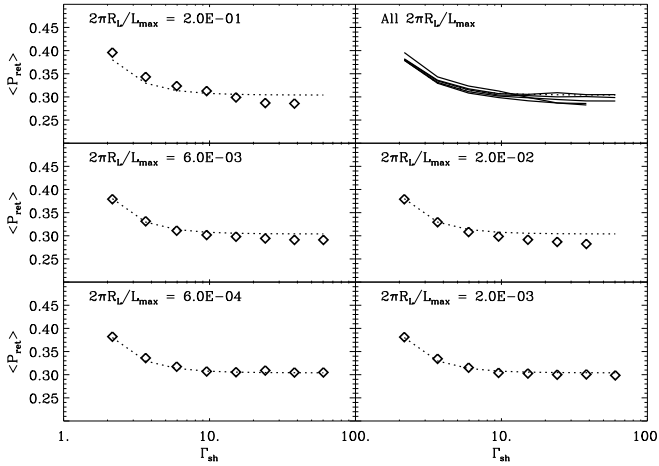}
\caption[]{Downstream return probability averaged over ingress angle
{\it vs} shock Lorentz factors for various rigidities as
indicated. The dotted line is an ad-hoc fit $\langle P_{\rm
ret}\rangle = 0.97 -0.67\beta_{\rm sh}$ which accounts well for the
dependence of the return probability with shock Lorentz factor at low
rigidities.}
\label{fig:Pret_Gamma}
\end{figure}

  One may further average the return probability $P_{\rm ret, d}$ over
the ingress pitch angle in order to define the average return
probability $P_{\rm ret}$:

\begin{equation}
\langle P_{\rm ret}\rangle\,\equiv\,{\int{\rm d}\mu^{\rm i}_{\rm d}\,
P_{\rm ret, d}(\mu^{\rm i}_{\rm d})\over \int{\rm d}\mu^{\rm i}_{\rm
d}}.
\label{eq:avpret}
\end{equation}
This probability is shown as a function of shock Lorentz factor for
varying values of the rigidity $\rho$ in
Fig.~\ref{fig:Pret_Gamma}. The dotted line represents the empirical
fit: $\langle P_{\rm ret}\rangle \simeq 0.97 - 0.66\beta_{\rm sh}$
which provides a good approximation at low rigidities.  This figure
shows how the average return probability reaches an asymptote with
$\Gamma_{\rm sh}$ for sufficiently low rigidities. At high rigidities
(upper panels), one recovers a dependence of $\langle P_{\rm
ret}\rangle$ on $\Gamma_{\rm sh}$. This latter effect is likely
related to the factor $R$ difference between $\rho_\parallel$ and
$\rho$: as $\rho_\parallel=R\rho$ becomes larger than $\approx 0.1-1$,
particles can no longer interact resonantly with the turbulence wave
modes (Casse {\it et al.} 2002); they take a longer time to return to
the shock, see Section~3.3, and their return probability becomes
sensitive to the nature of the turbulence, hence to $\Gamma_{\rm sh}$.

\begin{figure}
\centering \includegraphics[width=0.5\textwidth,clip=true]{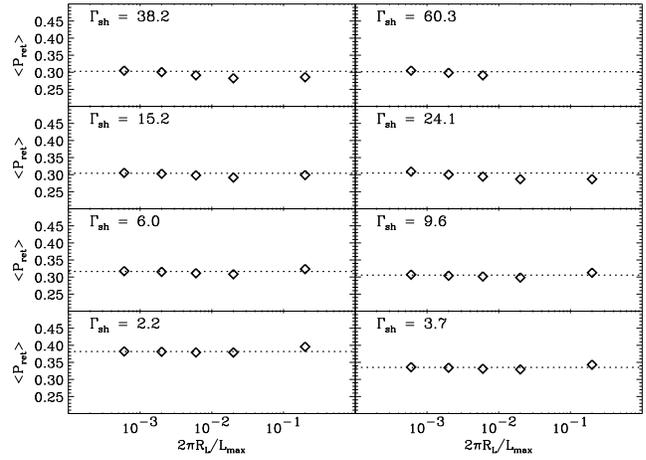}
\caption[]{Downstream return probability averaged over ingress angle
{\it vs} rigidity for various shock Lorentz factors, as indicated.}
\label{fig:Pret_E}
\end{figure}

  Finally one can plot the average return probability as a function of
rigidity for various values of the shock Lorentz factor, see
Fig.~\ref{fig:Pret_E}. It is important to note that the average return
probability does not depend on the rigidity, at least for sufficiently
low rigidities $\rho_\parallel\ll1$ for the same reasons as above. The
conditional return probabilities (from which $\langle P_{\rm
ret}\rangle$ is obtained) are also found not to depend on rigidity in
that range. In order to measure the spectral index of the accelerated
spectrum for rigidities in the inertial range of resonance, it is
important to use only the datasets of the smallest rigidities
downstream, i.e. $\rho=6\cdot 10^{-4}$ and $\rho=2\cdot 10^{-3}$,
where we still have $k_{\rm max}R_{\rm L}>1$, for the same reasons as
discussed above. For upstream probability laws, one can use all
datasets since the rigidities are well in the inertial range in the
absence of compression effects. In what follows, we use different
combinations of one downstream with one upstream of these datasets to
simulate the Fermi acceleration process and measure the spectral
index. We use these different datasets as independent realizations of
the conditional probability laws in order to estimate the numerical
uncertainty on the spectral index.

\subsection{Escaping accelerated particles}

 The fraction of particles that do not return to the shock adds up to
form the outgoing accelerated particle spectrum. As shown by Bell
(1978) for the case of non-relativistic shocks, the spectral index of
this spectrum is determined by the average return probability and the
mean energy gain at each cycle. For relativistic shocks, the
analytical development of Vietri (2002), whose formulation is very
similar to that presented in Section~2, shows that the spectral index
is determined by the energy gain properly averaged over the
equilibrium distribution functions both upstream and downstream, see
also Lemoine \& Pelletier (2003).

\begin{figure}
\centering \includegraphics[width=0.5\textwidth,clip=true]{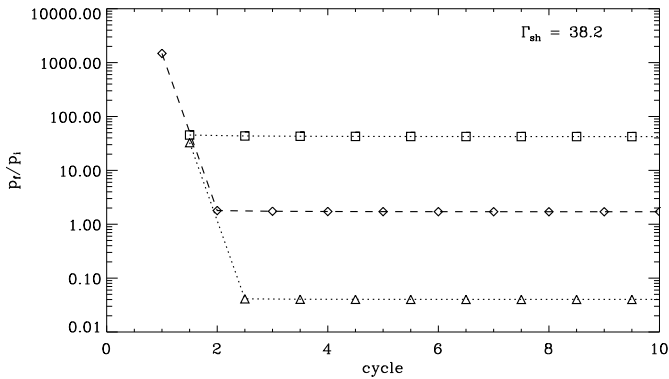}
\caption[]{Average energy gain per cycle u$\rightarrow$d$\rightarrow$u 
(diamonds), per half cycle u$\rightarrow$d (triangles), and per half 
cycle d$\rightarrow$u (squares) plotted vs successive cycles.}
\label{fig:cycle}
\end{figure}

  The average energy gains per cycle u$\rightarrow$d$\rightarrow$u and
half-cycles d$\rightarrow$u, u$\rightarrow$d are shown in
Fig.~\ref{fig:cycle}, which shows clearly that the gain is of order
$\simeq \Gamma_{\rm sh}^2$ for the first complete cycle
u$\rightarrow$d$\rightarrow$u, and falls to $\la 2$ in subsequent
cycles, as anticipated by Gallant \& Achterberg (1999), Achterberg
{\it et al.} (2001). This strong limitation of the energy gain is due
to the anisotropy of the distribution function upstream: particles do
not have time to be deflected by an angle greater than $\sim
1/\Gamma_{\rm sh}$ upstream before being overtaken by the shock which
moves at speed $\beta_{\rm sh}\simeq 1$ with respect to upstream,
hence the particles energy is decreased by a factor $1/\Gamma_{\rm
rel}$ in the half-cycle u$\rightarrow$d through the Lorentz transform,
in agreement with Fig.~\ref{fig:cycle}. In order to return to the
shock downstream, particles must turn back and the average energy gain
in the half-cycle d$\rightarrow$u is now $\simeq \kappa\Gamma_{\rm
rel}$, with $\kappa\la 2$ resulting in the total energy gain per cycle
$\la 2$. In the first cycle, the energy gain is large as the particle
population injected upstream toward the shock is isotropic, hence in
the first half-cycle u$\rightarrow$d the energy gain $\sim \Gamma_{\rm
rel}$.

  The average energy gain per cycle u$\rightarrow$d$\rightarrow$u is
$\langle p_{\rm f}/p_{\rm i}\rangle\simeq 1.7$ to within $\pm 0.1$ for
the various values of rigidity and shock Lorentz factor; this gain
tends to diminish with increasing $\Gamma_{\rm sh}$ albeit with a weak
slope.  A similar behavior has been observed in the case of isotropic
downstream turbulence (Lemoine \& Pelletier 2003), the gain decreasing
from $\simeq 2.0$ at $\Gamma_{\rm sh}=2$ to $\simeq 1.9$ at
$\Gamma_{\rm sh}\gg1$.

\begin{figure}
\centering \includegraphics[width=0.5\textwidth,clip=true]{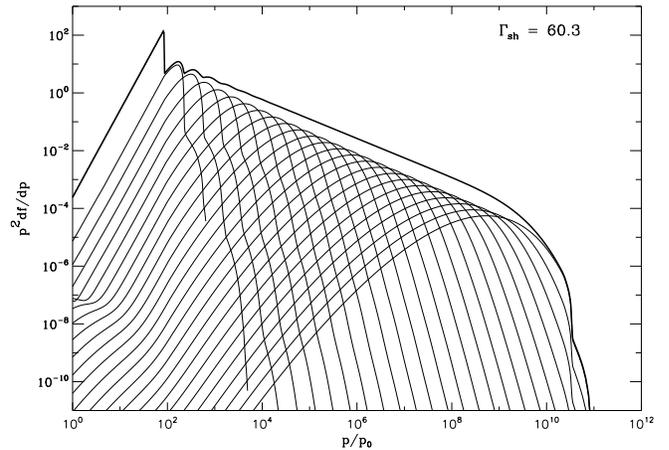}
\caption[]{Accelerated spectrum of particles escaping downstream times
momentum squared {\it vs} momentum (thick solid line); in thin solid
lines, the accelerated populations that escape downstream at each
cycle. }
\label{fig:spectra}
\end{figure}

Finally, using the method described in the previous section and the
probability data collected during the Monte Carlo simulations, one can
simulate the acceleration process itself and constructs the
accelerated particle population. The result is presented in
Fig.~\ref{fig:spectra}. This figure reveals that the sub-populations
that escape at each cycle $2n+1$, and whose spectrum is roughly a
gaussian centered on an energy $\sqrt{2}\Gamma_{\rm sh}p_0
g_{\text{u}\rightarrow \text{d}\rightarrow \text{u}}^n$ ($p_0$
injection energy) and amplitude $\propto (1-\langle P_{\rm
ret}\rangle)^n$, add up to form a featureless power law spectrum of
index $s$ (at $p\gg p_0$).

\begin{figure}
\centering \includegraphics[width=0.5\textwidth,clip=true]{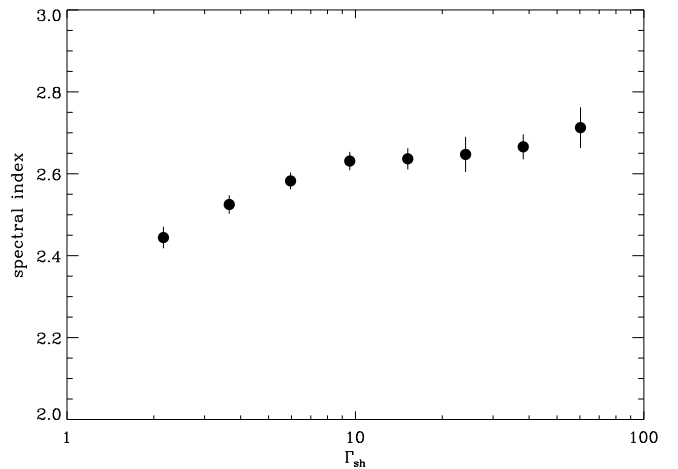}
\caption[]{Spectral index {\it vs} shock Lorentz factor.}
\label{fig:index}
\end{figure}

The measured spectral index $s$ is shown as a function of shock
Lorentz factor in Fig.~\ref{fig:index}. The comparison of these
results with those obtained for isotropic scattering downstream shows
that the inclusion of shock compression leads to a steeper accelerated
spectrum at all values of $\Gamma_{\rm sh}$. One can understand this
by noting that the compressed turbulence leads to lower average return
probabilities and slightly lower energy gains than those obtained for
isotropic turbulence (see Lemoine \& Pelletier 2003).

  The present results do not settle whether the spectral index reaches
an asymptote at large shock Lorentz factors, but at the very least, as
$\Gamma_{\rm sh}\gg 1$ it appears to evolve very weakly close to a
value $s\simeq 2.6-2.7$.

\subsection{Acceleration timescale}

The present simulations provide a direct measurement of the
acceleration timescale $t_{\rm acc}(\epsilon)$ at energy $\epsilon$,
which is defined as the u$\rightarrow$d$\rightarrow$u cycle timescale
in the upstream rest frame divided by the mean energy gain: 
\begin{equation}
t_{\rm acc}(\epsilon)\approx {t_{\text{u}\vert \text{u}}(\epsilon) +
\Gamma_{\rm sh}t_{\text{d}\vert \text{d}}(\epsilon/\Gamma_{\rm sh})\over 
g_{\text{u}\rightarrow \text{d}\rightarrow \text{u}}},
\end{equation} 
where $t_{\text{u}\vert \text{u}}$ and $t_{\text{d}\vert
\text{d}}$ are the upstream and downstream return timescales measured
in their respective rest frames.

\begin{figure}
\centering \includegraphics[width=0.5\textwidth,clip=true]{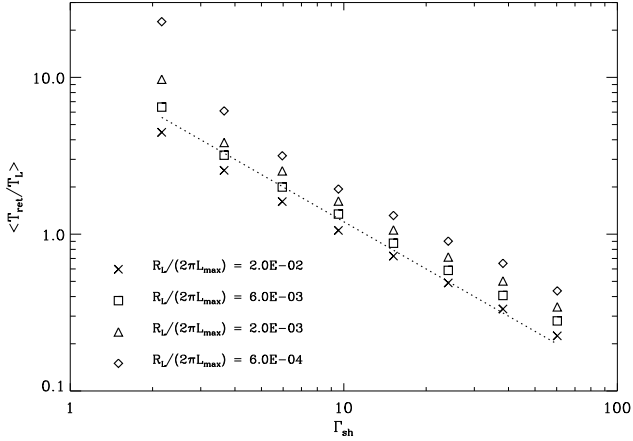}
\caption[]{Upstream return timescale in units of Larmor time averaged
over angular distribution {\it vs} shock Lorentz factor, for various
rigidities as indicated. The dotted line $T_{\rm ret}=12 T_{\rm
L}/\Gamma_{\rm sh}$ is shown as a guide the eye.}
\label{fig:tret_u_Gamma}
\end{figure}

  The upstream return timescale $t_{\text{u}\vert \text{u}}\sim
10t_{\rm L, u}/\Gamma_{\rm sh}$ ($t_{\rm L,u}$ upstream Larmor time)
up to a weak residual dependency on the rigidity, as shown in
Fig.~\ref{fig:tret_u_Gamma}. A fit that is accurate to a few percent
over the range of rigidities and for $\Gamma_{\rm sh}\ga 5$, is:
$t_{\text{u}\vert \text{u}}\simeq 14t_{\rm L,u}\rho^{0.19}/\Gamma_{\rm
sh}^{0.85}$. These results agree with and confirm the expectations of
Gallant \& Achterberg (1999) and Achterberg {\it et al.} (2001) who
argued that $t_{\text{u}\vert \text{u}}\propto 1/\Gamma_{\rm sh}$
since the particles are promptly overtaken by the shock when they have
been deflected by an angle of order $1/\Gamma_{\rm sh}$.

\begin{figure}
\centering \includegraphics[width=0.5\textwidth,clip=true]{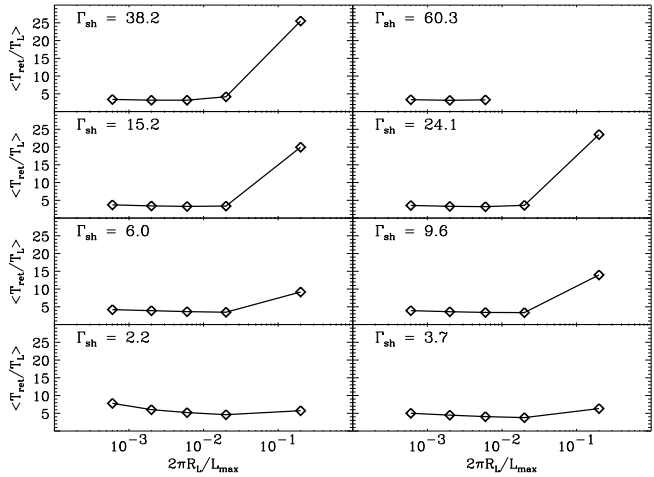}
\caption[]{Downstream return timescale in units of Larmor time
averaged over angular distribution {\it vs} rigidity for various shock
Lorentz factors as indicated.}
\label{fig:tret_d_E}
\end{figure}

  The downstream return timescale is plotted {\it vs} rigidity for
various shock Lorentz factors in Fig.~\ref{fig:tret_d_E}. This figure
shows that the return timescale $t_{\rm d\vert d}\approx 3-4 R_{\rm
L}/c$ at low rigidities $\rho\ll 0.1$ and $\Gamma_{\rm sh}\gg 1$. The
uncertainty in the numerical prefactor contains a weak residual
dependence on the shock Lorentz factor $t_{\rm d\vert d}\propto
\Gamma_{\rm sh}^{-0.08}$ ($\Gamma_{\rm sh}\gg1$).  Note that for the
moderately relativistic shock $\Gamma_{\rm sh}=2.2$, the return
timescale also contains a weak dependence on rigidity, $t_{\rm d\vert
d}\propto \rho^{-0.13}$ approximately, which disappears at larger
shock Lorentz factors. Here as well one can interpret the behavior of
$t_{\rm d\vert d}$ as the result of reflections of particles on the
compressed turbulence: the first scattering takes place on a Larmor
timescale and flings the particles back to the shock with probability
$\approx 0.3$; if the particle does not return to the shock after this
first scattering, the probability of doing so at subsequent
scatterings becomes negligible for two reasons: the shock moves away
at high velocity $\approx c/3$, and the enhancement of the tangential
components of the magnetic field prohibits efficient transport along
the shock normal.

  At high rigidities, the downstream return timescale increases; the
increase is all the more pronounced as the shock Lorentz factor is
high. This is related to the difference between $\rho_\parallel$ and
$\rho$; as $\Gamma_{\rm sh}$ increases, at a fixed value $\rho\sim
0.1$, the effective $\rho_\parallel$ becomes larger than $0.1-1$ and
the particles leave the range where resonant scattering with
turbulence modes along the shock normal is possible.  The scattering
time thus increases, see Section 4.1, and so does the return
timescale.

  Finally, the downstream return timescale can be written as $t_{\rm
d\vert d}(\epsilon/\Gamma_{\rm sh})\simeq 4R_{\rm L,
d}(\epsilon/\Gamma_{\rm sh})/c\simeq \sqrt{3}R_{\rm L,
u}(\epsilon)/\Gamma_{\rm sh}^2$, since the magnetic field strength is
amplified by $\sqrt{2/3}R$. Hence the downstream return timescale
measured in the upstream rest frame is slightly smaller than the
upstream return timescale, and the total acceleration time: $t_{\rm
acc}\approx 10 R_{\rm L\vert u}/\Gamma_{\rm
sh}$. Interestingly, the acceleration timescale becomes a fraction of
a Larmor time at large shock Lorentz factor, which may allow
acceleration to an energy limited by confinement arguments, in
particular $R_{\rm L}\la L_{\rm max}$, rather than by energy losses.

\section{Discussion}

\subsection{Relaxation length and transport in anisotropic turbulence}

The present study has hitherto assumed that the downstream turbulence
is successfully described by the direct compression of the upstream
turbulence through the shock jump conditions. However one must expect
this anisotropic compressed turbulence downstream to relax on a
timescale $\tau_{\rm rel}$ (defined in the downstream rest frame),
hence on a length scale $l_{\rm rel}=\beta_{\rm sh\vert d}c\tau_{\rm
rel}$ downstream. Particles will then experience this compressed
turbulence during their journey downstream provided the average
distance traveled from the shock front $l_{\rm tr}\sim t_{\rm d\vert d}/2\la
l_{\rm rel}$, or
\begin{equation}
t_{\rm d\vert d}\la 2\beta_{\rm sh\vert d}\tau_{\rm rel}.
\label{eq:relax}
\end{equation}
Otherwise the particles reach the point where the turbulence
anisotropy has relaxed and the previous considerations do not hold;
however, as we now argue, the above inequality is generally satisfied
in relativistic shocks for which the magnetic field is dynamically
unimportant.

  As discussed in Section 4.3, the return timescale $t_{\rm d\vert
d}\approx 3-4 R_{\rm L}/c$ for $\Gamma_{\rm sh}\gg1$ and $2\pi R_{\rm
L}/L_{\rm max}\ll 1$. There is a residual powerlaw dependence on both
rigidity and shock Lorentz factor but whose power indices are $< 0.1$,
see Section 3.3, which we can neglect for now. A simple but somewhat
na\"{\i}ve estimate for the (scale dependent) relaxation timescale is
$\tau_{\rm rel}\sim (k v_{\rm A})^{-1}$, where $k=2\pi/l$ is the eddy
wavenumber (related to the eddy size $l$), and $v_{\rm A}$ is the
Alfv\'en velocity. Since particles of Larmor radius $R_{\rm L}$
diffuse through resonant interactions with turbulent modes of
wavenumber $k\approx 1/R_{\rm L}$, the effective relaxation timescale
to be considered is $\tau_{\rm rel}\sim R_{\rm L}/v_{\rm A}$. The
inequality Eq.~\ref{eq:relax} is thus satisfied when $v_{\rm A}/c \la
0.2$ (for $\Gamma_{\rm sh}\gg 1$), which agrees with the hypothesis
made in Section~2 that the magnetic field is dynamically unimportant.
Interestingly, one may show that the bound on $v_{\rm A}$ is more
stringent for non-relativistic shocks, since the return timescale (for
isotropic scattering at least) scales as $t_{\rm d\vert d}\propto
t_{\rm scatt}/\beta_{\rm sh\vert d}$ in that case.

  The estimate for $\tau_{\rm rel}$ is likely to be conservative since
the eddy turn over rate, which gives a refined estimate of the
relaxation time on a scale $k$, reads: $\tau_{\rm t-o}\approx
(kv_k)^{-1}$, where $v_k$ is now the turbulent velocity on the scale
$k$, which decreases with increasing $k$; for a Kolmogorov spectrum,
$v_k\propto k^{-5/3}$. With this new estimate $\tau_{\rm rel}\sim
(kv_{\rm A})^{-1}(kL_{\rm max}/2\pi)^{5/3}$, the previous condition on
$t_{\rm ret}$ reads: $\rho\la 0.02(\beta_{\rm sh\vert d}/v_{\rm
A})^3$. Since $R_{\rm L}\sim L_{\rm max}/2\pi$ marks the maximal
energy reached in all likelihood, due to loss of confinement for $\rho
> 1$, inequality Eq.~\ref{eq:relax} is valid at all rigidities if
$v_{\rm A}\la 0.1 c$. Particles that are accelerated at the shock wave
thus do not travel far enough downstream to see anything else than the
turbulence in its compressed state.

\begin{figure}
\centering \includegraphics[width=0.5\textwidth,clip=true]{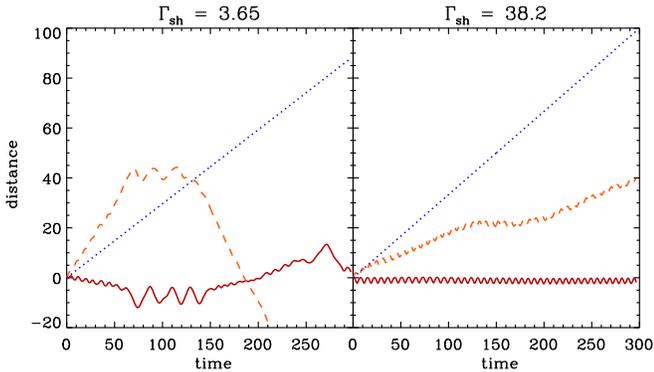}
\caption[]{Typical trajectory for a particle propagating downstream in
shock compressed turbulence with shock Lorentz factor $\Gamma_{\rm
sh}=3.7$ (left) and $\Gamma_{\rm sh}=38$ (right). The dotted line
indicates the trajectory of the shock front; the solid line shows the
displacement along the shock normal while the dashed line gives the
displacement in the plane parallel to the shock front.}
\label{fig:traj}
\end{figure}

  The transport of particles in strongly compressed turbulence is
peculiar, as illustrated by Fig.~\ref{fig:traj}. It presents examples
of particle trajectories downstream for two different values of the
shock Lorentz factor but for the same upstream magnetic configuration;
in both cases the particle never returns to the shock. The comparison
of the typical displacement along and perpendicular to the shock
normal indicates that the particles appear confined in a layer of
turbulence that lies tangential to the shock plane, and for periods of
time extending well beyond a Larmor timescale. Note that for
$\Gamma_{\rm sh}=38$ (right panel), the particle appears to gyrate
along a magnetic field line located in a plane parallel to the shock
front. In this particular case, the magnetic field configuration is
locally transverse. If the trajectory is followed for a sufficiently
long period of time, it will depart from a simple Larmor gyration
law. Moreover the effective pulsation is not constant along the
trajectory shown in Fig.~\ref{fig:traj}, but varies slightly in a
random way. It is also apparent in these figures that the
characteristic pitch angle scattering timescale along the shock normal
is of order of a Larmor time, while that measured in the perpendicular
direction is much larger. This demonstrates qualitatively how
particles return to the shock in a few Larmor times by reflecting on
the compressed turbulent modes.

\begin{figure}
\centering \includegraphics[width=0.5\textwidth,clip=true]{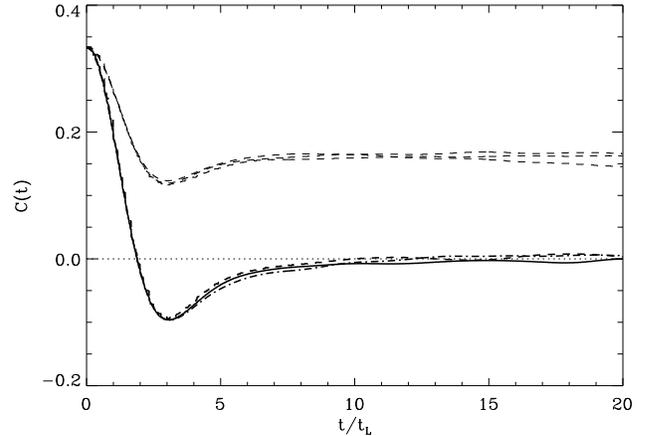}
\caption[]{Velocity correlation function $C(\tau)$ {\it vs.}  time (in
units of Larmor time $t_{\rm L}$) for various values of rigidity
$\rho$ and shock Lorentz factor $\Gamma_{\rm sh}$. The lower thick
curves correspond to the velocity oriented along the shock normal; in
thick solid line, $\rho=3\cdot10^{-6}$ and $\Gamma_{\rm sh}=3.65$; in
thick dashed line, $\rho=3\cdot10^{-6}$ and $\Gamma_{\rm sh}=38.2$; in
thick dashed-dotted line, $\rho=10^{-3}$ and $\Gamma_{\rm
sh}=3.65$. The upper dotted lines show the correlation function for
the velocity components perpendicular to the shock normal; at
$\tau\geq15$, the upper two are $\rho=3\cdot10^{-6}$ and $\Gamma_{\rm
sh}=3.65,38.2$, the lower curve is for $\rho=10^{-3}$ and $\Gamma_{\rm
sh}=3.65$.}
\label{fig:corr}
\end{figure}

  In order to better characterize the transport of particles in
compressed turbulence, one may seek the diffusion coefficients in the
various directions. For our purposes, it is more relevant to study the
time correlation function of the particle velocities, $C_{ij}(\tau)$:
\begin{equation}
C_{ij}(\tau)\,=\,\langle v_i(\tau) v_j(0)\rangle,\label{eq:corr}
\end{equation}
where the average is to be taken on a large number of trajectories,
and $v_i(\tau)$ is the velocity of the particle in the direction $i$
at time $\tau$. The integration of $C_{ij}(\tau)$ over $\tau$ leads to
the diffusion coefficient $D_{ij}$ (Candia \& Roulet 2004).  In the
present case, the correlation function is however more relevant since
particles never actually diffuse downstream before returning to the
relativistically moving shock. The correlation functions along the
shock normal, $C_{\parallel}=C_{zz}$ and perpendicular to the shock
normal, $C_{\perp}=(C_{xx}+C_{yy})/2$ are shown for various values of
$\Gamma_{\rm sh}$ and $\rho$ in Fig~\ref{fig:corr}. This figure
demonstrates that the early time behavior of the parallel correlation
function is, to a high degree of accuracy, independent of both
$\Gamma_{\rm sh}$ and $\rho$, provided $\Gamma_{\rm sh}\gg 1$ and
$\rho\ll1$. The time behavior of $C_\parallel$ can be grossly
approximated by $C_\parallel(\tau)\sim \cos(2\pi \tau/t_{\rm
L})\exp(-\tau/\tau_\parallel)$, and $\tau_\parallel$ gives the
scattering time along the shock normal. This fit is not reproduced on
Fig.~\ref{fig:corr} for the sake of clarity, and because it diverges
from the measured curves for $\tau/t_{\rm L}\ga 10$. However, one can
see by eye that the estimated $\tau_\parallel \approx 3t_{\rm L}$.

  On the contrary the velocities perpendicular to the shock normal do
not decorrelate on timescales as short as $\tau_\parallel$; the
fall-off of $C_\perp(\tau)$ toward zero is observed (but not shown on
Fig.~\ref{fig:corr}) on much longer timescales than $10-20t_{\rm L}$,
and the decorrelation time $\tau_\perp$ is found to depend on
$\rho$. This is expected insofar as the scattering time in isotropic
Kolmogorov turbulence $\tau_\perp/t_{\rm L}\propto \rho^{-2/3}$ (Casse
{\it et al.} 2002). Indeed, in Fig.~\ref{fig:corr}, the lower dashed
curve corresponds to the high rigidity value, and it is seen to fall
off more rapidly than the other two (that correspond to a same
rigidity) at $\tau/t_{\rm L}\geq 15$. One may also note the slight
decorrelation in perpendicular velocities induced at early times by
the decorrelation of velocities along the shock normal.

  This figure thus nicely explains the transport properties that were
indirectly observed in the previous discussion, namely that the
scattering time along the shock normal, which is the relevant quantity
for shock acceleration, is of order of three Larmor times, and
independent of rigidity and shock Lorentz factor to a good
approximation.

\subsection{Comparison to previous results}

To our knowledge, there is no existing study of Fermi acceleration in
ultra-relativistic shock waves which includes the effect of
compression of the upstream magnetic field. One may nevertheless find
interesting points of comparison in various limits with studies by
Ballard \& Heavens (1992), Ostrowski (1993), Bednarz \& Ostrowski
(1998), Kirk {\it et al.} (2000) and Niemiec \& Ostrowski (2004).

 Ballard \& Heavens (1992) were the first to attempt modeling of Fermi
acceleration in non-relativistic to moderately relativistic
($\Gamma_{\rm sh}\la 5$) shocks with shock compressed turbulence by
the means of Monte Carlo methods. They found a pronounced steepening
of the spectra index with increasing shock speed and derived the
approximate formula $s\simeq 0.75\Gamma_{\rm sh}+1.25$. Although our
study confirms the increase of $s$ with increasing values of
$\Gamma_{\rm sh}$, the precise value of $s$ differs significantly from
those of Ballard \& Heavens (1992), all the more so at large shock
velocities. One may probably attribute this discrepancy to the modest
dynamical range ($64-100$) that was available at the time of the
simulations of Ballard \& Heavens (1992). If the dynamic range is not
large enough, the particle rigidity always lie close to the upper
range of resonance and this results in steeper spectra due to
increased escape probability. The subsequent study by Ostrowski (1993)
obtained much harder spectra than Ballard \& Heavens (1992) where
suitable comparison can be made. In particular, for large turbulence
amplitude $\delta B/B =3$, and shock Lorentz factors $\Gamma_{\rm
sh}=(2.3,5.0)$, Ostrowski (1993) obtained $s=(2.0-2.2,2.2-2.3)$, where
the range of values bracket different values of the mean field
inclination with respect to the shock normal. Our results for $s$
indicate slightly larger values for $s$, but the agreement is
generally better than with Ballard \& Heavens (1992).

Kirk {\it et al.} (2000) have studied relativistic Fermi acceleration
using semi-analytical eigenfunction methods; their results confirm the
canonical value $s=2.23$ in the case of isotropic scattering. They
have attempted to address the effect of anisotropic scattering
downstream using an analytical description of compressed turbulence
and an analytical estimate of the diffusion coefficient. They
concluded that anisotropy does not affect significantly $s$, a result
which is clearly at odds with the present study. The source of the
discrepancy lies probably in the modeling of downstream diffusion by
Kirk {\it et al.} (2000): as should be obvious from
Figs.~\ref{fig:corr}, particles propagating in strongly compressed
turbulence downstream do not actually diffuse downstream but rather
turn back by reflecting on a compressed magnetic layer.

 The most detailed study to date is that of Niemiec \& Ostrowski
(2004), who have studied Fermi acceleration in moderately relativistic
shocks ($\Gamma_{\rm sh}\la 5$) by Monte Carlo integration of the
particle trajectory in a magnetic field with a large dynamic
range. Among the results obtained, the authors quote a generic
non-power law behavior of the accelerated spectrum: the spectra
generally appear harder close to the cut-off ($\rho \sim 1$) than well
below the cut-off.  Niemiec \& Ostrowski (2004) conclude that this
effect is probably related to the finite dynamic range: close to the
cut-off, the propagation regime in compressed turbulence differs from
that in the inertial range of resonance, and one indeed finds a
different return probability or mean energy gain, both of which
control the value of $s$. Our Fig.~\ref{fig:spectra} reveals a smooth
powerlaw behavior at energies well beyond the injection point; this is
expected on the grounds that the conditional return probability
histograms used to model Fermi acceleration is itself rigidity
independent. We have demonstrated in the previous Sections that, deep
in the inertial range $\rho \ll 0.1$, this is a good approximation;
however, we have also observed that as $\rho$ tends to larger values,
the return probability reveals a slight dependence on $\rho$, and this
would make $s$ evolve with $\rho$, albeit for $\rho \ga
10^{-2}-10^{-1}$, had we included this dependence in our calculations.

  Our results disagree markedly from those of Niemiec \& Ostrowski
(2004) with regards to the rigidity dependence of the return
probability in the inertial range. These authors claim to observe a
pronounced non-monotonic rigidity dependence, which is definitely
absent from our simulations down to the percent level. We note that
the noise level in the results of Niemiec \& Ostrowski (2004) is not
indicated in the figures, and might account for part of this apparent
variability. On theoretical grounds, there is neither expectation nor
justification for a rigidity dependence of $\langle P_{\rm
ret}\rangle$ on $\rho$, as long as resonant interactions with the
turbulence can occur and the shock lifetime is infinite. There is no
clear explanation or interpretation of this observed rigidity
dependence of the return probability in Niemiec \& Ostrowski
(2004). Furthermore their measured value of $\langle P_{\rm
ret}\rangle$ does not agree with ours: for $\Gamma_{\rm sh}\simeq5$
and $\delta B/B=3$, they find $\langle P_{\rm ret}\rangle \simeq
0.20$, significantly lower than ours. This results in a steeper
spectrum with $s\simeq 2.9\pm0.1$, to be compared with our value
$s\simeq2.6$; one should note that Niemiec \& Ostrowski (2004) warrant
caution with respect to their analysis of $\Gamma_{\rm sh}=5$, as it
lies close to the limits of their simulation capabilities.

  We also note that the results of Niemiec \& Ostrowski (2004) differ
signicantly from those of Ostrowski (1993), although the method used
is similar. For $\beta_{\rm sh}=0.5$, $\delta B/B=3$ and mean field
inclination $\Psi=45^o$, Niemiec \& Ostrowski (2004) find $s=2.7$
while Ostrowski (1993) obtains $s=2.0\pm 0.1$ for the same values (but
$\Psi=50^o$). This difference persists at larger inclinations, $s=2.8$
in the former {\it vs} $s=2.\pm0.1$ in the latter for $\Psi=75^o$,
other values unchanged; it also persists at larger shock velocities,
in particular for $\beta_{\rm sh}=0.9$ ($\Gamma_{\rm sh}=2.3$) and
$\Psi=45-50^o$, $s=2.5$ in the former {\it vs} $s=2.\pm 0.1$ in the
latter. Here as well the source of the discrepancy remains unknown.
Overall, and where comparison can be made, our results lie halfway
between those of Ostrowski (1993) and those of Niemiec \& Ostrowski
(2004).

   As stressed at the end of Section~2, the present method offers a
significant gain in signal when compared to direct Monte Carlo methods
but it cannot take into account the possible correlations between
upstream and downstream trajectories due to the correlations between
upstream and downstream magnetic fields at the point of
shock-crossing. This remark, when taken together with the above
comparison to previous work, shows the need for more exhaustive
studies of relativistic Fermi acceleration with shock compressed
turbulence, including mean magnetic fields of various strength and
obliquity, and using both the present method and direct Monte Carlo
methods.

\subsection{Applications to astrophysical shock waves}

Gamma$-$ray bursts, with their shock Lorentz factor in excess of 100
are ideal candidates to test theories on particle acceleration in
ultra-relativistic flows. In the standard fireball model (see Meszaros
2002, Piran 2004 for reviews), the prompt $\gamma-$emission is
interpreted as the product of synchrotron emission of electrons
accelerated in the internal shocks with Lorentz factors $\Gamma_{\rm
sh}\sim 2-5$ in the comoving wind frame. The afterglow emission is
interpreted as the synchrotron light of electrons accelerated in the
ultra-relativistic shock wave with $\Gamma_{\rm sh}\sim 300$, that
itself results from the interaction of the $\gamma-$ray ejecta with
its environment.

 The spectral indices of the shock accelerated electrons derived in
both cases are $s\simeq2.3$. This has been interpreted as a dramatic
confirmation of our understanding of relativistic Fermi acceleration
since it agrees with the canonical value $s=2.2$ obtained for
isotropic downstream scattering. However, as should be clear by now,
this ``agreement'' rather reflects our poor understanding of the
acceleration process: the inclusion of shock compressed turbulence,
which should be seen as a refinement of the theory, leads to steeper
spectra, see Section~4, with $s\simeq 2.4-2.6$ for $\Gamma_{\rm
sh}=2-5$, and $s\simeq 2.6-2.7$ for $\Gamma_{\rm sh}\gg 1$. 

  The difference is not as significant in the case of internal shocks
than for afterglow observations. As a matter of fact, isotropic
scattering downstream predicts a value $s\simeq 2.1-2.2$ for
$\Gamma_{\rm sh}=2-5$ (Lemoine \& Pelletier 2003), hence it could not
account reasonably well for the dispersion observed in the spectral
slopes of $\gamma-$ray prompt emission. However it is possible that
the inclusion of a mean magnetic field component with varying
inclinations and, possibly varying turbulence level, could reproduce
this dispersion. Moreover it is not yet established whether the
$\gamma$ radiation results from synchrotron emission of shock
accelerated electrons; other radiating processes (e.g. Piran 2004 and
references therein) or magnetic reconnection events in the flow (e.g.
Lyutikov \& Blandford 2003) are likely possibilities.

  Concerning the discrepancy of the present spectral index with that
inferred from afterglow observations, one must note that the present
study is limited to the case of pure Kolmogorov turbulence upstream,
which idealizes the limit $\delta B/B\gg1$. However, judging by the
comparison with Niemiec \& Ostrowski (2004), one does not expect the
inclusion of a coherent component to help resolve this discrepancy, as
these authors have observed a steepening of the accelerated spectrum
with decreasing turbulence level $\delta B/B$. In the
ultra-relativistic regime, this trend should be exacerbated, since as
$\delta B/B$ decreases, one approaches the perpendicular shock
acceleration limit where Fermi acceleration becomes inefficient
(Begelman \& Kirk 1990). Our simulations should thus provide a
conservative lower bound to $s$ for the case that includes a mean
magnetic field.

  The discrepancy might be attributed to the nature of the turbulence,
in particular to the assumption of Kolmogorov turbulence. Again, the
work of Niemiec \& Ostrowski (2004) suggests that the turbulence
spectral index has an effect on $s$, although there is not enough
simulation data to pinpoint what the exact correlation is. These
remarks indicate the need for more exhaustive studies that investigate
various turbulence spectra. Interestingly, this suggests that prompt
and afterglow observations of $\gamma-$ray bursts might be giving us
information on the properties of the turbulence behind the shock
front; the upstream turbulence does not play any r\^ole in the
ultra-relativistic limit as a result of the limited amount of time
that a particle spends upstream before being overtaken by the shock
front.

  The interpretation of $\gamma-$ray bursts afterglows as synchrotron
emission by shock accelerated electrons requires that the magnetic
field intensity at the shock front be significantly higher than the
average interstellar value (e.g. Piran 2004 and references
therein). The nature of downstream turbulence would then be directly
related to the amplification process, which might manifests itself
indirectly in the spectral slope. The proposal of magnetic field
amplification by the two-stream Weibel instability (Medvedev \& Loeb
1999) has recently triggered a lot of interest. This instability seems
able of explaining the high value of $B$ required, although debate on
the subject is not closed, see Wiersma \& Achterberg (2004). In any
case, the Weibel instability amplifies the magnetic field in the
transverse plane to the shock normal and on very small spatial scales
(Medvedev \& Loeb 1999) . If the magnetic field on scales larger than
or equal to the Larmor radius of a typical accelerated particle is
thus not amplified by the instability, one should expect acceleration
to proceed as presented here, and the discrepancy should remain.
These considerations may suggest that the amplified magnetic field
structure differs from that proposed by Medvedev \& Loeb (1999) or
that the amplification mechanism itself is different. In this regard,
we note that recent numerical studies on the Weibel instability
suggests that powerlaw acceleration may occur independently of the
Fermi mechanism due to the presence of electromagnetic currents
downstream (Hededal {\it et al.}  2004). It should also be noted that
stochastic acceleration in the downstream turbulence, which has not
been accounted for in the present study, could play a significant
r\^ole in reshaping the accelerated spectra, as suggested by Virtanen
\& Vainio (2005).

  On a different line of thought, one should point out that the value
of the spectral index derived here turns out to be in very good
agreement with that required to fit the ultra-high energy part of the
cosmic ray spectrum at energies $E\ga 10^{18}\,$eV, namely $s\simeq
2.6-2.7$, when one assumes that the sources are distributed at
cosmological distances and do not evolve too strongly with redshift
relatively to the cosmic star formation rate (Berezinsky {\it et al.}
2002, Berezinsky {\it et al.}  2005, Lemoine 2005).

\section{Summary}

  We have conducted a study of Fermi acceleration at relativistic and
ultra-relativistic shock waves, considering the effect of the shock
compression on the downstream magnetic turbulence. The numerical
simulations are based on Monte Carlo methods of particle propagation
in realistic magnetic fields described by sums of plane wave
modes. The numerical technique differs from the standard Monte Carlo
modeling of Fermi acceleration in that it measures the relevant
statistical laws of particle transport on either side of the shock,
and uses these probability laws together with the Lorentz transform
from one frame to the other to reconstruct the acceleration process.

  The turbulence was assumed to be described by pure Kolmogorov
turbulence upstream, a situation which idealizes the limit $\delta B/B
\gg1$. The main effect of the compression with respect to the case of
isotropic scattering is to steepen the accelerated spectrum to a slope
$s\simeq 2.6-2.7$ in the limit $\Gamma_{\rm sh}\gg1$, as a result of a
decreased return probability. This latter effect is induced by the
compression, which amplifies the magnetic field in the transverse
direction to the shock normal: particles that enter downstream are
trapped on a Larmor timescale in a compressed turbulence layer and
cannot recross the shock unless they turn back within a few Larmor
times.  Consequently, the acceleration timescale is dominated by the
usptream residence time, and can be as short as $t_{\rm acc}\simeq
10t_{\rm L}/\Gamma_{\rm sh}$ (upstream frame). We have also argued
that the accelerated particles do not travel far enough downstream
before returning to the shock to experience a turbulence that has
relaxed to near isotropicity.

  The derived slope does not agree with that inferred from
observations of $\gamma-$ray bursts afterglows, which indicate
$s\simeq2.3$. This inferred value is generally accepted as a success
of Fermi acceleration in the relativistic regime whose predicted
canonical value is $s\simeq 2.2-2.3$. However, this result only holds
for isotropic scattering downstream, whereas the inclusion of
realistic shock jump conditions, as done here, makes downstream
turbulence strongly anisotropic and the spectra markedly steeper. The
resolution of this discrepancy may be tied to the necessary but
unknown amplification mechanism of the upstream magnetic field.

\section*{Acknowledgments}

We thank A. Marcowith and G. Pelletier for useful discussions and
insight on the relaxation length of compressed turbulence.

\label{lastpage}

\end{document}